\documentclass[
   final,          
  ]
  {aipproc}

\usepackage{graphicx}
\usepackage{epsfig}

\layoutstyle{8x11single}

\begin{document}

\title{Random Quantum Circuits and Pseudo-Random Operators: Theory and Applications}

\author{Joseph Emerson}{
  address={Perimeter Institute for Theoretical Physics}
}

\begin{abstract}
Pseudo-random operators consist of sets of operators that exhibit
many of the important statistical features of uniformly
distributed random operators. Such pseudo-random sets of operators
are most useful whey they may be parameterized and generated on a
quantum processor in a way that requires exponentially fewer
resources than direct implementation of the uniformly random set.
Efficient pseudo-random operators can overcome the exponential
cost of random operators required for quantum communication tasks
such as super-dense coding of quantum states and approximately
secure quantum data-hiding, and enable efficient stochastic
methods for noise estimation on prototype quantum processors. This
paper summarizes some recently published work demonstrating a
random circuit method for the implementation of
\emph{pseudo-random} unitary operators on a quantum processor
[Emerson et al., Science 302:2098 (Dec.~19, 2003)], and further
elaborates the theory and applications of pseudo-random states and
operators.
\end{abstract}

\maketitle

\section{Introduction}

Random numbers play a fundamental role in classical information
theory, with practical applications including system
identification, stochastic simulation, and a variety of
cryptographic protocols. Similarly, random quantum states and
operators have both fundamental and practical importance to
quantum information theory. The operator ensemble with greatest
relevance to quantum information is the circular unitary ensemble
(CUE), defined as the distribution of unitary operators drawn
uniformly with respect to the unique, unitarily invariant (Haar)
measure on $U(D)$, where $D = 2^{n_q}$ is the dimension of Hilbert
space and $n_q$ is the number of qubits. The Haar measure induces
a uniform measure over pure states and thus a random unitary
matrix may be applied to randomize an arbitrary pure state. This
leads to a number of practical applications for quantum
communication: sets of randomizing unitaries enable the
super-dense coding of arbitrary quantum states
\cite{Harrow-Superdense}, lead to a decrease in the classical
communication cost for remote state preparation and allow the
construction of more efficient data hiding schemes
\cite{Bennett-RSP}, and provide a means to reduce the shared key
length for the (approximate) encryption of quantum states
\cite{Hayden-epsRandom}. The usefulness of these protocols is
limited by the fact that generating \emph{uniformly random}
operators and states requires quantum resources that grow
exponentially in the number of quantum bits and hence these
protocols become impractical for large quantum systems. Below I
describe the random circuit method for generating \emph{efficient}
sets of \emph{pseudo-randomly} distributed unitary operators that
was originally introduced in Ref.~\cite{Emerson03}, and then
describe how random and pseudo-random operators enable efficient
methods for noise-estimation on quantum processors.

\section{Random Circuits}

A unitary operator drawn from the Haar measure on $U(D)$ is
conveniently parameterized via the Hurwitz decomposition
\cite{Zyc2}. Using standard techniques this decomposition can be
re-expressed as a quantum circuit requiring  $O(\log(D)^2 D^2)$
elementary (one and two qubit) gates and $D^2$ independent random
`input' parameters, and therefore, as expected, requires resources
growing exponentially in the number of qubits.

To overcome this impracticality, we consider instead a random
circuit comprised of a sequence of $m$ iterations of a
constant-depth gate parameterized by independent random input
parameters. The constant-depth gate has two steps, as illustrated
in Figure 1. The first step of the gate consists of rotating each
qubit by independent, random unitary operators drawn from the Haar
measure on $U(2)$. This first step requires $3 n_q$ independent
input variables. The second step of the constant-depth gate
consists of the following set of simultaneous two-body
interactions, $U = \exp( i (\pi/4) \sum_{j=1}^{n_q-1} \sigma_z^j
\otimes \sigma_z^{j+1} )$, where $\sigma_z^j$ is the usual Pauli
operator of the $j$'th qubit and the coupling angle is fixed at
$\pi/4$ to maximize the entanglement produced by the two-body
interactions. Since we imagine a simple 1-D array of qubits and a
spatially local coupling interaction, we have limited our basic
circuit to include only nearest-neighbor couplings.

\begin{figure}
\includegraphics{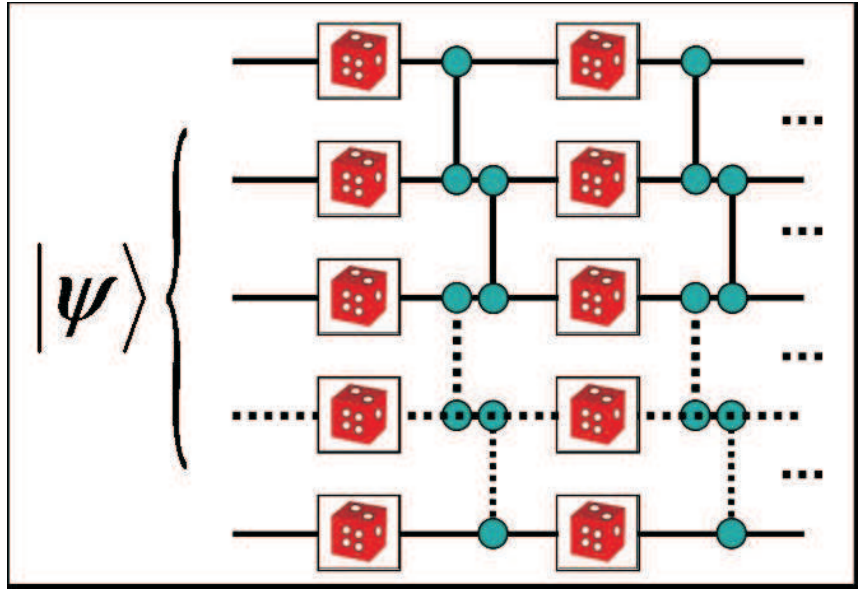}
\caption{A random circuit comprised of a sequence of iterations of
a constant-depth gate parameterized by independent random input
parameters. The constant-depth gate has two steps. The first step
of the gate consists of rotating each qubit by random unitary
operators drawn independently from the Haar measure on $U(2)$.
These random local rotations are indicated by the boxes containing
dice. The second step of the constant-depth gate consists of any
convenient nearest-neighbor coupling between the qubits. The set
of pseudo-random unitary operators generated by this method
converges asymptotically (in the strong sense of uniform
convergence) to the Haar measure on $U(D)$ \cite{Emerson04} and
can efficiently reproduce practical signatures of uniformly (Haar)
distributed random unitary operators
\cite{Emerson03,Cucchietti04}.}
\end{figure}

Our construction induces a measure over the set of circuits of
length $m$.  For finite $m$ the distribution of unitary operators
generated by the random circuit is of course biased with respect
to the uniform (Haar) measure on $U(D)$.  In this sense the random
circuits generate ``pseudo-random'' unitary operators. For $m >
m_c = O(n_q^3 D^2)$, this distribution, though biased, does enjoy
non-vanishing support over all elements of the group $U(D)$.
Moreover, in the limit of increasing numbers $m$ of iterations,
the measure over the composed circuits converges exponentially (in
the uniform sense) to the uniform measure on the group
\cite{Emerson04}. This result follows from properties of the
Fourier transform on compact groups and requires that the initial
distribution has non-vanishing support over a universal gate set.
It is worth stressing that this argument indicates that a random
circuit based on {\em any} universal gate set generates the
uniform measure asymptotically, no matter how biased the method
with which the gates are drawn from the universal set.

We now turn to the question of whether the measure over random
circuits of length $m <$ polylog($D$), i.e., efficient random
circuits, can mimic some practical subset of the statistical
features of the uniform measure. We are in particular interested
in properties of unitary operators and states that are relevant to
quantum information tasks, and so we first consider the
distribution of entanglement produced by the circuit.
Specifically, we consider the entanglement of the states obtained
by applying the random circuit to the computational basis states.
As a practical indicator of entanglement for multi-partite systems
we consider the indicator proposed by Meyer and Wallach
\cite{Meyer}, which can be expressed as an average of the
bi-partite entanglement between each qubit and the rest of the
system \cite{Brennen},
\begin{equation}
Q = 2 - \frac{2}{n_q}\sum_{i=1}^{n_q} \textrm{Tr}[\rho_i^2],
\label{MeyerEnt}
\end{equation}
where $\rho_i$ is the reduced density operator of the $i$'th
qubit. One has $0 \le Q \le 1 $, with $Q = 0$ for completely
factorable states and $Q=1$ for the cat states. In
Ref.~\cite{Emerson03} we report the distribution of $Q$ produced
by the random circuits, as a function of increasing $m$, for an 8
qubit system. The distribution becomes indistinguishable (on the
scale of the figure) from the distribution produced by CUE maps.
It is useful to note that the exact CUE average,
\begin{equation}
\langle Q \rangle = \frac{D-2}{D+1} \simeq 1 - \frac{3}{D} +
O(D^{-2}),
\end{equation}
approaches the maximum value of $Q$ exponentially as a function of
$n_q$. Moreover, the standard deviation decreases exponentially
with increasing $n_q$ \cite{Caves}.
Thus in the limit of large $n_q$ almost all random (and
pseudo-random) states have nearly maximal entanglement $Q \simeq
\langle Q\rangle \simeq Q_{max}$. This ``concentration of
measure'' effect is a typical feature of sufficiently smoothly
varying functions of states in Hilbert spaces of large dimension
\cite{Haake,Mehta,Hayden-genEnt}. Broadly speaking, a single
complex system may be represented by a generic pure state in a
large Hilbert space. Generic pure states are ``typical'' or
``universal'' in the sense that many of their properties are well
approximated by the value of that property obtained by averaging
with respect to the uniform measure over all states.  Indeed it is
this universality feature that has allowed a wide variety of
complex quantum systems to be characterized by the universal and
analytically tractable averages calculated with respect to the
relevant ensemble.

While the above results and arguments indicate that pseudo-random
unitaries can mimic the uniformly distributed random unitaries for
the smoothly varying properties of interest, it is crucial to
verify that they can do so efficiently.  As shown in the inset of
Fig. 2 in Ref.~\cite{Emerson03}, the rate at which the average $Q$
for the random circuits approaches the CUE average is indeed
exponential. However, it should be noted that as $n_q$ increases
the exponential convergence rate decreases. The rate of decrease
is itself decreasing, strongly suggesting the possibility that at
worst only polylog($D$) resources are required to converge to the
CUE average. Indeed preliminary results from subsequent work
suggest that the rate actually saturates at a finite asymptotic
value \cite{Cucchietti04}, indicating that the random circuit
method is indeed efficient, at least for the entanglement
indicator we have considered.

Another important feature of random unitary operators is the
marginal distribution of their matrix elements. This distribution
plays a key practical role in the approximate randomization of
quantum states
\cite{Harrow-Superdense,Hayden-epsRandom,Bennett-RSP}. Numerical
results in Ref.~\cite{Emerson03} indicate that the distribution of
matrix elements of the set of random circuits converges to the
distribution expected for the uniform measure at an exponential
rate. Similar results were obtained for the distribution of
eigenvector components. This latter feature has particular
relevance to quantum processing since it is the randomness of the
eigenvectors that enables methods for the unbiased estimation of
unknown noise sources \cite{Emerson02} via the algorithm described
below.

\section{Stochastic Methods for Noise Estimation}

Circuit implementations of random and pseudo-random unitaries
enable novel methods for efficiently characterizing the strength
and type of noise sources in imperfect quantum processors. As is
well-known, identification and correction of the dominant
coherent, incoherent and decoherent noise sources for a given
quantum device is an essential step toward the eventual
realization of fault-tolerant computation. The distribution and
strength of these noise operators will differ for different
devices and, moreover, the effect of the noise generators will
generally depend on the applied control fields
\cite{Pravia,Boulant}. The direct approach to this problem is
quantum process tomography, which scales inefficiently and becomes
impossible to implement in practice for devices with more than
just a few qubits \cite{Yaakov}. However, when the target
transformation is sufficiently random, some of the measurable
signatures of noise and decoherence, such as the rate of the
average fidelity decay \cite{Jacquod,Emerson02} and the average
rate of purity loss \cite{Paz2}, become independent of any system
properties and depend only on the intrinsic properties of the
noise \cite{Emerson02}. Moreover, by implementing a
motion-reversal algorithm, where a pseudo-random unitary operator
is applied $n$ times followed by its inverse $n$ times, the rate
of fidelity decay and purity loss arising from any unknown noise
sources may be measured via efficient algorithms \cite{Emerson02}
and subsequently compared to the analytic (ensemble average)
predictions (which depend only on the characteristics of the noise
model). It should be noted that an exponentially large (complete
set) of initial states are not required (as in process tomography)
since almost all initial states lead to fidelity decay rates that
are exponentially close to the ensemble average
\cite{Blume-Kohout04}. In this way the implementation of
pseudo-random unitary operators on a quantum processor enables
\emph{efficient stochastic} algorithms for the unbiased
measurement of intrinsic properties of the unknown noise sources.

\section{Discussion}

While numerical evidence indicates that random circuits can
efficiently reproduce many of the statistical features of the
uniform measure that are relevant to quantum information tasks,
rigorous proofs of efficiency and error bounds are still needed.
This is especially important in the case of the quantum
cryptography applications, where a proper security analysis is
required. In the context of noise estimation, new techniques are
needed that can determine a wider set of features of the noise,
and in particular those features that are most relevant to
assessing the scalability of a device as a platform for
fault-tolerant quantum computation, and to optimizing the
appropriate choice of active and passive error-correction
protocols.

\begin{theacknowledgments}
This paper is based on a talk given at QCMC04. The random circuit
method described in this paper was developed in collaboration with
Y. Weinstein, M. Saraceno, S. Lloyd, and D. Cory and published
originally in Ref.~\cite{Emerson03}. That work (and many of the
still unpublished results cited above) benefitted from helpful
discussions with R. Blume-Kohout, F. Cucchietti, J. Goldstone, A.
Harrow, E. Livine, P. Zanardi, and K. Zyczkowski.
\end{theacknowledgments}

\end{document}